\documentclass[prc,twocolumn,showpacs,floatfix,nofootinbib,preprintnumbers,superscriptaddress,amsmath,amssymb]{revtex4-1}

\usepackage{graphicx}
\usepackage{epsfig}
\usepackage{amsmath,amssymb}
\usepackage{bm}
\usepackage{color}
\usepackage{float}
\usepackage{dcolumn} 

\newcommand{\pr}[1]{{\sc{\lowercase{#1}}}}
\graphicspath{ {images/} }

\begin{document}

\title{Third minima in thorium and uranium isotopes in the self-consistent theory}

\author{J.D. McDonnell}
\affiliation{Department of Physics \&
Astronomy, University of Tennessee, Knoxville, Tennessee 37996, USA}
\affiliation{Physics Division, Oak Ridge National Laboratory, Oak Ridge, Tennessee 37831, USA}
\affiliation{Physics Division, Lawrence Livermore National Laboratory, Livermore, California 94551, USA}

\author{W. Nazarewicz}
\affiliation{Department of Physics \&
Astronomy, University of Tennessee, Knoxville, Tennessee 37996, USA}
\affiliation{Physics Division, Oak Ridge National Laboratory, Oak Ridge, Tennessee 37831, USA}
\affiliation{Institute of Theoretical Physics, University of Warsaw, ul. Ho\.za
69, 00-681 Warsaw, Poland}

\author{J.A. Sheikh}
\affiliation{Department of Physics \&
Astronomy, University of Tennessee, Knoxville, Tennessee 37996, USA}
\affiliation{Physics Division, Oak Ridge National Laboratory, Oak Ridge, Tennessee 37831, USA}
\affiliation{Department of Physics, University of Kashmir, Srinagar, 190 006, India}

\date{\today}

\begin{abstract} 
\begin{description}
 \item[Background]
Well-developed third minima, corresponding to strongly elongated and reflection-asymmetric shapes associated with dimolecular configurations, have been predicted in some non-self-consistent models to impact fission pathways of thorium and uranium isotopes.  These predictions have guided the interpretation of resonances seen experimentally.  On the other hand, self-consistent calculations consistently predict very shallow potential-energy surfaces in the third minimum region. 
\item[Purpose]
We  investigate the interpretation of third-minimum configurations in terms of dimolecular (cluster) states.
We study the isentropic potential-energy surfaces of selected even-even thorium and uranium isotopes at several excitation energies.  In order to understand the driving effects behind the presence of third minima, we study the interplay between pairing and shell effects. 
\item[Methods]
We use the finite-temperature superfluid nuclear density functional theory. We consider two Skyrme energy density functionals: a traditional functional SkM$^*$ and a recent functional UNEDF1  optimized for fission studies.  
\item[Results]
We  predict very shallow or no third minima in the potential-energy surfaces of $^{232}$Th and $^{232}$U.
In the lighter Th and U isotopes with $N=136$ and 138, the third minima are better developed. 
We show that the reflection-asymmetric configurations around the third minimum can be associated with dimolecular states involving the spherical doubly magic $^{132}$Sn and a lighter deformed Zr or Mo fragment.
The potential-energy surfaces for $^{228, 232}$Th and $^{232}$U at several excitation energies are presented.  We also study isotopic chains to demonstrate the evolution of the depth of the third minimum with neutron number.  
\item[Conclusions]
We show that the neutron shell effect that governs the existence of the dimolecular states around the third minimum is consistent with  the spherical-to-deformed shape transition in the Zr and Mo isotopes around $N=58$. 
We demonstrate that the depth of the third minimum is sensitive to the excitation energy of the nucleus. In particular, the thermal reduction of pairing, and related enhancement of shell effects, at small excitation energies help to develop deeper third minima. At large excitation energies, shell effects are washed out and third minima disappear altogether. 
\end{description}

\end{abstract}

\pacs{24.75.+i, 21.60.Jz, 27.90.+b, 24.10.Pa}

\maketitle

\section{Introduction}\label{section:th232intro}

The phenomenon of nuclear fission is a large-amplitude collective motion in which the nucleus undergoes a series of shape rearrangements  before splitting into distinct fragments.  
The observables for a fissioning system, such as fission half-life and properties of fission fragments, are sensitive to the sequence of nuclear configurations through which the nucleus is driven on the way to fission \cite{BjornholmRev52,Wagemans,KrappeFissionBook}. Local minima in the potential-energy surface, often  representing metastable configurations, can profoundly affect the dynamics and timescale of fission. Of particular importance are 
superdeformed fission isomers \cite{metag1980,SinghNDS2002}, corresponding to the ``second mininum" in actinide nuclei, separating  inner and outer saddles.  Another important class of states consists of hyperdeformed ``third minima" predicted theoretically in the early seventies \cite{Pashkevich1971,Moller1972} and soon afterwards attributed to the resonance microstructures observed in the fission cross sections found in the light actinides
\cite{Britt72,*Blons78}. Continued experimental studies of the actinides \cite{BlonsNPA414,BlonsNPA502,BelliaPRC20,Baumann1989,NenoffEPJA32,Piessens1993,BlokhinPAN2009,Krasznahorkay1998,KrasznahorkayPLB1999,CsigePRC80,Csatlos2005,CsigePRC85} inferred the existence of highly elongated minima, and its reflection asymmetric structure has been supported by the presence of parity doublets
\cite{ThirolfPPNP49}. 

The appearance of third minima around $^{232}$Th has been attributed to large shell effects associated with reflection asymmetric configurations
corresponding to dimolecular structures \cite{NazDob1992,Aberg1994,Shneidman2000,Royer2006},  with one fragment resembling doubly magic $^{132}$Sn \cite{CwiokPLB322,TerAkopan1996,PashkIJMPE18}.
Pronounced  third minima have been predicted in theoretical studies  of thorium and uranium isotopes, especially those carried out with the macroscopic-microscopic (MM) approach \cite{Pashkevich1971,Moller1972,MollerIAEA,Bengtsson1987,CwiokPLB322,TerAkopan1996}. On the other hand,
self-consistent  studies based on the nuclear density functional theory \cite{RutzNPA590,BonneauEPJA21,BergerNPA502,Delaroche2006103,McDonnellFissionProc2009,*McDonnellPhD2012}, as well as recent MM work  \cite{KowalPRC85,*Jachimowicz:2013fk}, typically find a third minimum that is much shallower than that of the earlier MM calculations or the empirical barrier fits \cite{CapoteNDS110,SinPRC74}. This result is 
 puzzling in light of of the accumulated  experimental  evidence (resonances in fission cross sections, mass and kinetic-energy distributions of fission fragments, fits to experimental cross sections, moments of inertia, and presence of parity doublets).  

To clarify the situation, we carry out self-consistent calculations 
for eight even-even Th and U isotopes  
within the superfluid, finite-temperature nuclear density functional theory  (FT-DFT), investigating how the potential-energy surfaces and the third minima evolve with  excitation energy. In particular,
we seek to isolate the contributions to the nuclear energy that may be responsible for third minima. 
We review the FT-DFT model in Sec.~\ref{section:model}.   Section~\ref{section:th232isochains} presents an analysis of the trends seen in potential-energy curves and shell correction energies across Th and U isotopic chains, finding that deep third minima appear in lighter isotopes. 
To study the interpretation of third-minimum configurations in terms of dimolecular states, we analyze total nucleonic densities.
We proceed to analyze two-dimensional, finite-temperature potential-energy surfaces of $^{228,232}$Th and $^{232}$U in Sec.~\ref{section:th232pes}.  As the excitation energy increases and pairing quenches, we actually find a regime in which the third minimum is slightly deepened for $^{232}$Th.  Finally, the conclusions of our work are given in Sec.~\ref{section:th232conclusions}.  

\section{The Model}\label{section:model}

\begin{figure}[htb]
 \includegraphics[width=1\columnwidth]{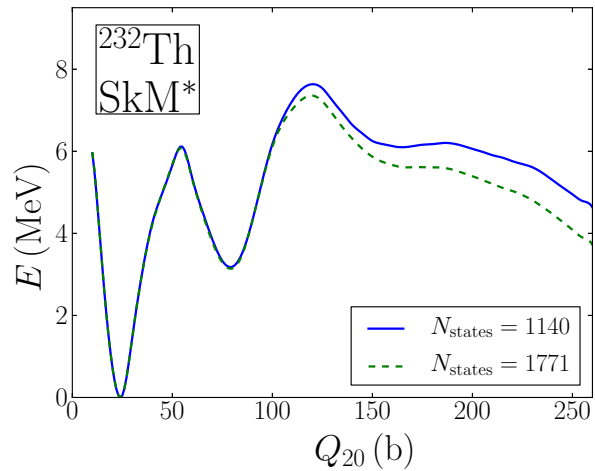}
 \caption[T]{\label{fig:th232_basis}
 (Color online) Potential-energy curve for $^{232}$Th with SkM$^*$ obtained with a basis of 1140 (solid line) and 1771 (dashed line) stretched HO basis states.  }
\end{figure}

To study the potential-energy surfaces (PESs) as a function of the excitation energy $E^*$, we employ the superfluid FT-DFT theory \cite{GoodmanNPA352,EgidoNPA451,PeiPRL102} in the implementation of Refs.~\cite{SheikhPRC80,McDonnellFissionProc2009,*McDonnellPhD2012}. We employ the 
symmetry-unrestricted Skyrme DFT code  \pr{HFODD} \cite{DobCPC180,SchunckCPC2012}, which solves the  the  finite-temperature  
Hartree-Fock-Bogoliubov (HFB) 
equations in the Cartesian harmonic oscillator (HO) basis.  The oscillator length is varied according to the method of Refs.~\cite{Dobaczewski:1997dq,Staszczak:2005lq,KortelainenPRC85}.  
This basis choice is a compromise between accuracy and time of calculation that has been studied and used successfully in the past (see Ref.~\cite{Staszczak:2005lq} and Fig.~6 of Ref~\cite{NikolovPRC83}).  The work of Ref.~\cite{Schunck:2012nx} estimates that the use of $1000$ to $1200$ HO basis states can produce an error up to $2$ to $3$\,MeV beyond the second fission barrier.  For $^{232}$Th, we compare a calculation of the SkM$^*$ PES with 1140  and 1771 HO basis states in Fig.~\ref{fig:th232_basis}.  We do see that including more basis states reduces the absolute value of the potential energy beyond the second barrier by about $0.5$\,MeV, but that the topology of the PES is hardly affected.  Therefore, in our two-dimensional and finite-temperature PESs, we chose to utilize the basis of the lowest 1140 stretched HO  basis states originating from $31$ major oscillator shells.  

To constrain the total quadrupole moment $Q_{20}$ (elongation) and total octupole moment $Q_{30}$ (reflection asymmetry, important at outer saddle and on to scission) we employ  the augmented Lagrangian method \cite{StasEPJA2010}.  We use a mesh with step sizes of five units in each collective degree of freedom --- for smooth PESs, we interpolate with second-order splines.  We also apply a constraint on the triaxial quadrupole moment $Q_{22}$  to force the system to break axial symmetry, subsequently relaxing this constraint to allow the system to follow the minimum-energy path in the $Q_{22}$ direction.  Consequently, the inner fission barrier heights are lowered due to the axial symmetry breaking. 
As discussed in  previous papers \cite{Staszczak09,WardaPRC86,Staszczak2013}, exploring many collective coordinates and associated symmetry breaking enables us to identify saddle points and static valleys \cite{Moll09all,Dubray12} as the competing adiabatic fission pathways are well separated in the collective space.

The finite-temperature HFB equations are obtained from the minimization of the grand canonical potential, so that the free energy $F = E - TS$ is formally calculated at a fixed temperature $T$. While the fission process is not isothermal, it is reasonable to treat the collective motion during fission as an adiabatic process \cite{NazarewiczNPA1993}.  We thus assume that the entropy is constant during this adiabatic motion, and we exploit the correspondence between surfaces of free energy at constant temperature and surfaces of internal energy at constant entropy \cite{Diebel1981,Faber1984}. This equivalence, based on   Maxwell's relations,  has  been verified numerically in the self-consistent calculations of Ref. \cite{PeiPRL102}.

We map the excitation energy of the nucleus $E^*$ to the fixed temperature $T$ via
\begin{equation}
E^* (T) = E_{\mathrm{g.s.}}(T) - E_{\mathrm{g.s.}}(T = 0),
\end{equation}
where $E_{\mathrm{g.s.}}(T)$ is the minimum energy of the nucleus at temperature $T$.  This corresponds well to the excitation energy of a compound nucleus \cite{PeiPRL102,SheikhPRC80}.

To study the role of shell effects in producing the third minimum, we use the the Strutinsky energy theorem \cite{RingSchuckBook} to decompose the self-consistent energy $E$:
\begin{equation}
 E = E_{\mathrm{smooth}} + \delta E^{\mathrm{sh}}, 
\end{equation}
where $E_{\mathrm{smooth}}$ is a bulk contribution to the energy that varies smoothly with nucleon number
and $\delta E^{\mathrm{sh}}$ is a shell correction energy. To extract
$\delta E^{\mathrm{sh}}$ from the HFB energy, we employ the procedure described in
Refs.~\cite{VertsePRC61,NikolovPRC83} with the smoothing width parameters
$\gamma_n = 1.54,  \gamma_p = 1.66$ (in units of $\hbar\omega_0 = 41/A^{1/3}$ MeV) and the curvature correction $p = 10$.

The nuclear interaction in the particle-hole channel has been approximated through the SkM$^*$ parametrization \cite{BartelNPA386} of the Skyrme energy density functional (EDF).  This traditional  EDF achieves realistic surface properties in the actinides, allowing a good description of the evolution of the energy with deformation \cite{Staszczak09,WardaPRC86,Staszczak2013}.  In the particle-particle channel, we use the density-dependent mixed-pairing interaction \cite{Dobaczewski02}. Our calculations with SkM$^*$ at each excitation energy were performed at the HFB level 
with a quasiparticle cutoff energy of $E_{\rm cut} = 60$\,MeV. The pairing strengths
$V_{\tau 0}$ ($\tau=n, p$) are chosen to fit the pairing gaps determined from experimental odd-even mass differences in $^{232}$Th \cite{AME2003}.  For SkM$^*$ EDF, the pairing strengths are $V_{n0} = -273.5$\,MeV and
$V_{p0} = -334.0$\,MeV.

At $E^* = 0$\,MeV, we also performed calculations with the recently developed EDF parametrization UNEDF1   \cite{KortelainenPRC85}.  In  UNEDF1 calculations, 
we restore approximately the
particle number symmetry broken in HFB by using the variant of the Lipkin-Nogami scheme with the cutoff energy $E_{\rm cut} = 60$\,MeV \cite{Stoitsov03}.  
Because the UNEDF1 functional relies on the Lipkin-Nogami treatment of pairing, 
and the  corresponding
finite-temperature HFB Lipkin-Nogami extension has not been implemented,  the free-energy results presented in this paper are based on SkM$^*$.

\section{Third minima in T\lowercase{h} and U isotopes}\label{section:th232isochains}
\begin{figure}[htb]
 \includegraphics[width=0.8\columnwidth]{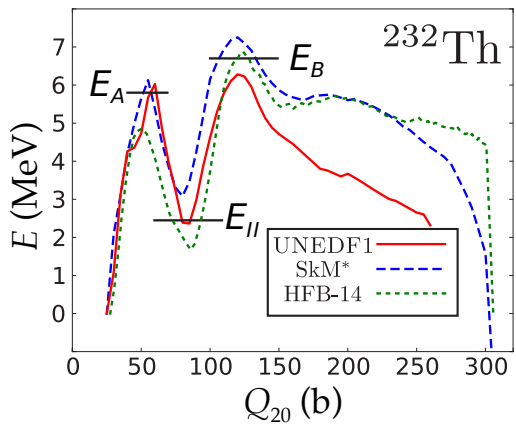}
 \caption[T]{\label{fig:th232_func_compare}
 (Color online) Potential-energy curves for $^{232}$Th obtained with  several EDFs: UNEDF1 and SkM$^*$ (this work), and HFB-14 \cite{GorielyPRC75}. The empirically inferred values of the first and second barrier heights $E_A$ and $E_B$ \cite{CapoteNDS110}, as well as the measured energy of the fission isomer $E_{II}$ \cite{BrowneNDS107}, are marked.
}
\end{figure}
Self-consistent calculations  tend to predict either  a very shallow or no third minimum  for $^{232}$Th and $^{232}$U.  As seen in Fig.~\ref{fig:th232_func_compare}, 
the potential-energy curves for $^{232}$Th, obtained with several EDFs, each exhibit a gentle downwards slope beyond the second saddle ($E_B$) --- none of these models predicts a large third hump.  A shallow third minimum appears around $Q_{20}=165$\,b in  our SkM$^*$ model and $Q_{20}=150$\,b in the HFB-14 calculations of Ref.~\cite{GorielyPRC75}. This minimum seems to be more pronounced in the relativistic DFT calculations of
Ref.~\cite{RutzNPA590} employing PL-40, NL1, and NL-SH functionals.
Our UNEDF1 results show a local plateau at $Q_{20} \approx 200$\,b, but the third barrier is practically nonexistent.

The accuracy of any statement about the existence of the third minimum relies on the use of a sufficiently large HO basis.  Indeed, previous calculations with D1S \cite{BergerNPA502} exhibited a shallow third minimum in $^{232}$Th. With a larger basis \cite{PhysRevLett.85.30,*WardaComm}, however, this minimum flattens into a plateau similar to that seen in the SkM$^*$ and HFB-14 calculations of Fig.~\ref{fig:th232_func_compare}.  As discussed in Sec.~\ref{section:model},  in this work we use a sufficiently large HO basis so that the final results  are not sensitive to basis choice.

The results shown in Fig.~\ref{fig:th232_func_compare} and Ref.~\cite{RutzNPA590} indicate that self-consistent models  predict a shallow third minimum, or a softness in the PES of $^{232}$Th, in the region beyond the outer saddle with  $Q_{20} \approx 150 - 200$\,b. This indicates that
the shell effects responsible for this structure  are systematically present in DFT calculations, but their strength is strongly model dependent.  To identify the model features that are most conducive to third minima, we turn to focus on our calculations with SkM$^*$ and UNEDF1.

\begin{figure}[htb]
 \includegraphics[width=1\columnwidth]{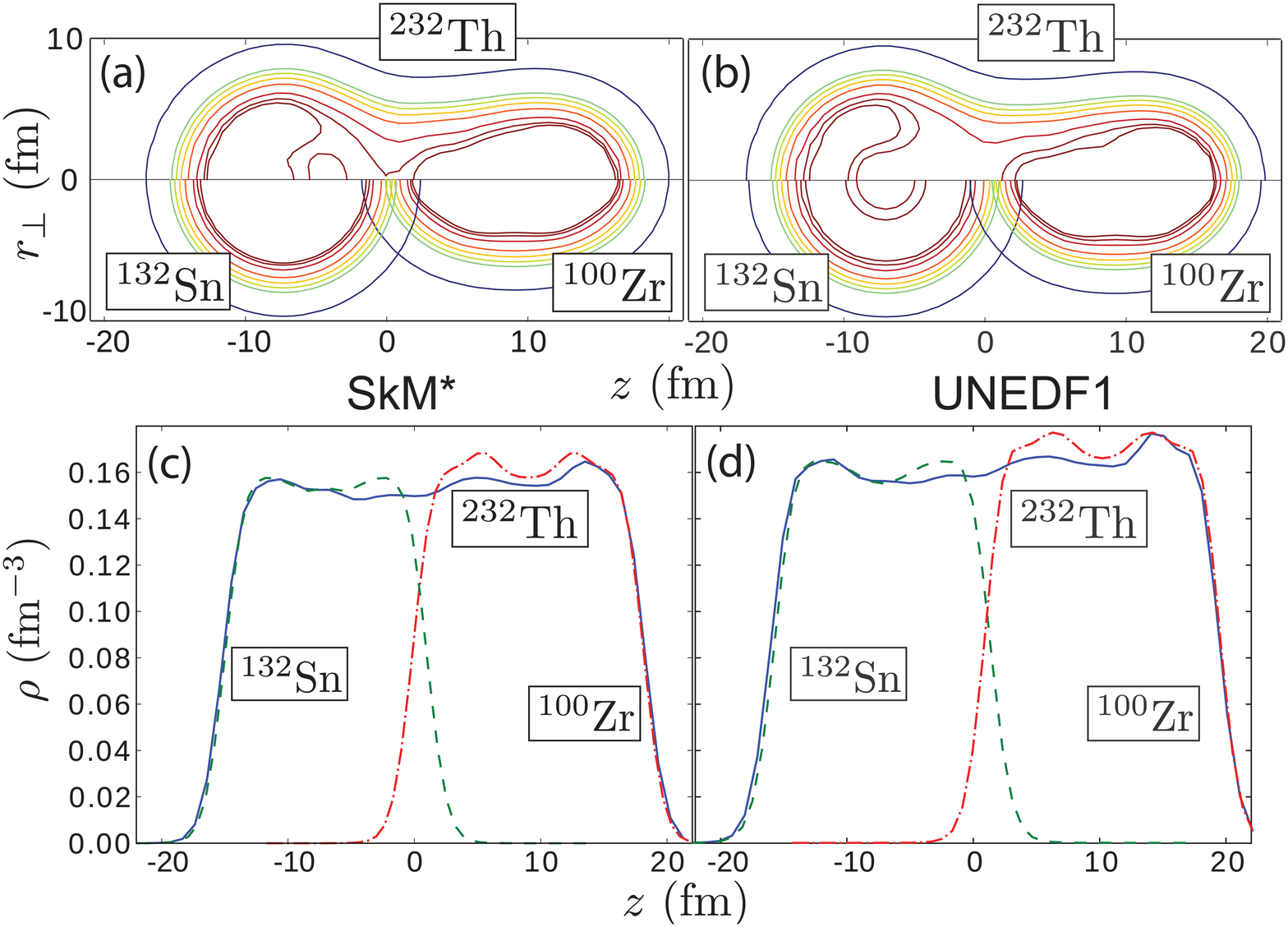}
 \caption[T]{\label{fig:th232_density}
 (Color online) (top) Cross section    of total density of $^{232}$Th in $yz$ plane calculated with (a) SkM$^*$ and (b) UNEDF1 at third minimum ($Q_{20}=165$\,b),   compared to cross sections of $^{132}$Sn and $^{100}$Zr densities. (bottom) Density profiles for $^{232}$Th and the $^{132}$Sn and $^{100}$Zr  fragments along the $z$ axis obtained in (c) SkM$^*$ and (d) UNEDF1.}
\end{figure}

The third minimum of $^{232}$Th has been associated with a dimolecular configuration, in which one fragment bears a strong resemblance to the doubly-magic $^{132}$Sn \cite{CwiokPLB322,TerAkopan1996}.  The nuclear density profiles of $^{232}$Th, $^{132}$Sn, and $^{100}$Zr, calculated with SkM$^*$ and UNEDF1, are displayed and compared in Fig.~\ref{fig:th232_density}. (In making this comparison, we followed the methodology of Ref.~\cite{WardaPRC86}.) Namely, the configuration of
$^{232}$Th corresponds to  $Q_{20}=165$\,b, the configuration of $^{132}$Sn corresponds to its spherical ground state, and 
the $^{100}$Zr fragment configuration corresponds to its prolate ground state with $Q_{20}=10$\,b. 
The resemblance of the left-hand fragment of $^{232}$Th to $^{132}$Sn is clearly seen in both models, although the nascent fragments overlap to produce the sizable neck seen in Fig.~\ref{fig:th232_density}. 

As discussed in, e.g.,  Refs.~\cite{NenoffEPJA32,Piessens1993,TerAkopan1996},
the high likelihood of obtaining $^{132}$Sn-like fragments in the fission of actinides can be attributed to the doubly-magic nature of  $^{132}$Sn.  The recent theoretical studies of the asymmetric fission around $^{180}$Hg  \cite{WardaPRC86, PanebiancoPRC86} indicate that the shell effects at prescission configurations associated with the deformed fragment also play a  significant role in the determination of  fission yields.

\begin{figure}[htb]
 \includegraphics[width=\columnwidth]{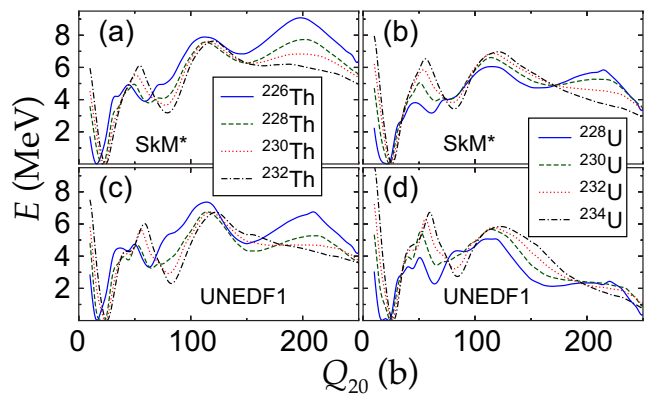}
 \caption[T]{\label{fig:E8Total}
  (Color online) Potential-energy curves predicted for (left) $^{226,228,230,232}$Th and (right) $^{228,230,232,234}$U with
  (top) SkM$^*$ and (bottom) UNEDF1 EDFs.  
}
\end{figure}

A more comprehensive  survey with SkM$^*$ and UNEDF1 shown in Fig.~\ref{fig:E8Total} reveals that the lighter isotopes of thorium and uranium,  $^{226,228}$Th and $^{228}$U, are expected to have deeper third minima. For 
$^{230}$Th and $^{230}$U, our SkM$^*$ calculations exhibit a shallow third minimum, which vanishes in UNEDF1. 

\begin{figure}[htb]
\includegraphics[width=0.8\columnwidth]{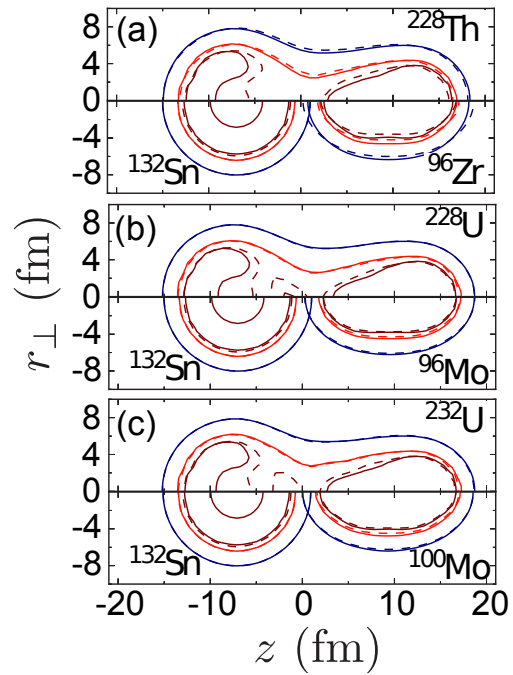}
\caption[T]{\label{fig:thu_densities}
(Color online) Contour plots for total densities of (a) $^{228}$Th,
  (b) $^{228}$U, and (c) $^{232}$U calculated with  UNEDF1 (solid lines) and SkM$^*$ (dashed lines) compared with fragment densities: spherical  $^{132}$Sn and a lighter deformed nucleus around $^{100}$Zr.  
 The contour levels shown are at $50\%$, $90\%$, and $95\%$ of the saturation density ($\rho_0 = 0.16$\,fm$^{-1}$).  }
\end{figure}

Is there evidence that third minima are exactly correlated with a dimolecular clustering in the density?  In Figs. \ref{fig:th232_density} (top) and \ref{fig:thu_densities}, we compare the density profiles of isotopes that exhibit third minima in our SkM$^*$ and UNEDF1 calculations
($^{228}$Th,  $^{228}$U)  
with those that do not ($^{232}$Th,  $^{232}$U).  In fact,  there is evidence for dimolecular clustering in each case.  It is interesting that both $^{228}$Th and $^{228}$U show slightly smaller necks than $^{232}$Th and $^{232}$U, respectively.  But the differences between the density profiles are not very dramatic -- isotopes that are predicted to have  third minima do not show significantly more dimolecular clustering than isotopes in which third minima are absent.

\begin{figure}[htb]
\includegraphics[width=0.8\columnwidth]{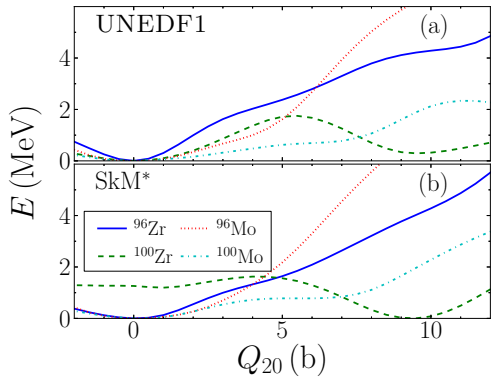}
\caption[T]{\label{fig:zr_totals}
(Color online) Potential-energy curves for $^{96,100}$Zr and $^{96,100}$Mo, calculated with (a) UNEDF1 and (b) SkM$^*$.    }
\end{figure}

For each of the cases presented, the heavier fragment is the spherical $^{132}$Sn while the lighter fragment is in a deformed configuration with $Q_{20} = 10$\,b.  As seen in Fig.~\ref{fig:zr_totals}, except for $^{100}$Zr, each of these nuclei is spherical in its ground state -- it is not reasonable to argue that a third minimum can be associated with nascent lighter fragments close to the nuclide's ground state.  However, it is interesting to note
that in all the nuclei discussed there is a competition between spherical and deformed configurations~\cite{Wood92,HeydeWood,Reinhard99}. For instance, $^{96}$Zr is believed to be spherical, but it has a deformed excited 0$^+$ state \cite{Wood92}.
The ground state  of $^{100}$Zr is  strongly deformed, with the coexisting spherical configuration lying  higher in energy. The  balance between
relative position of spherical and deformed configurations around $^{98}$Zr primarily depends on the size of  predicted $Z=40$ and $N=56$ single-particle gaps that vary from model to model~\cite{Reinhard99}. The differences between
SkM$^*$ and UNEDF1 predictions seen in Fig.~\ref{fig:zr_totals}
are thus indicative of subtle differences between the shell effects, which also play out to result in a shallower third-minimum region  in  UNEDF1.  

Why does the dimolecular configuration result in a deeper third minimum in lighter nuclei such as $^{228}$Th, and not in $^{232}$Th or $^{232}$U?  And why are the third minima present in UNEDF1 shallower than those of SkM$^*$?  The density profiles shown for UNEDF1 and SkM$^*$ in Figs. \ref{fig:th232_density} and \ref{fig:thu_densities} indicate that the densities predicted by
UNEDF1 and SkM$^*$ are in fact very similar, and the isotopic dependence is weak. We seek an answer in the underlying shell effects. 
\begin{figure}[htb]
 \includegraphics[width=\columnwidth]{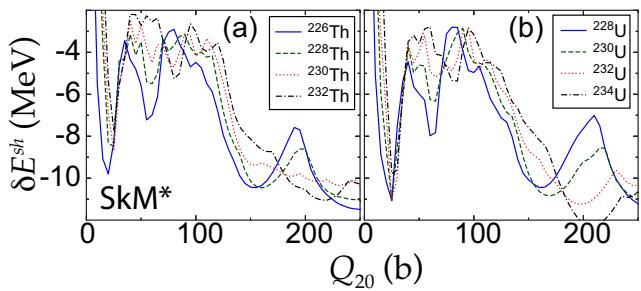}
 \caption[T]{\label{fig:TotalShell}
  (Color online) Total shell energies from SkM$^*$ for (a) $^{226,228,230,232}$Th and (b) $^{228,230,232,234}$U.
}
\end{figure}

The total shell energies calculated with  SkM$^*$ are displayed in Fig.~\ref{fig:TotalShell}. (The shell corrections obtained with UNEDF1 have a similar pattern but they are reduced in magnitude; hence, they are not shown.)
The shell corrections for the $N=136, 138$ isotopes, $^{226,228}$Th  and $^{228,230}$U,  indicate a 
strong shell effect at $Q_{20}\approx 150$\,b. For the  $N=140, 142$ isotopes ($^{230,232}$Th  and $^{232,234}$U), $\delta E^{\mathrm{sh}}$ tends to stabilize more elongated configurations, at $Q_{20}\approx 200$\,b.  
This result is reminiscent of a spherical-to-deformed shape transition around $N=58$ in the Zr and Mo isotopes discussed above -- associated with the lighter fragments in the  dimolecular picture of the third minimum.

 \begin{figure}[htb]
 \includegraphics[width=\columnwidth]{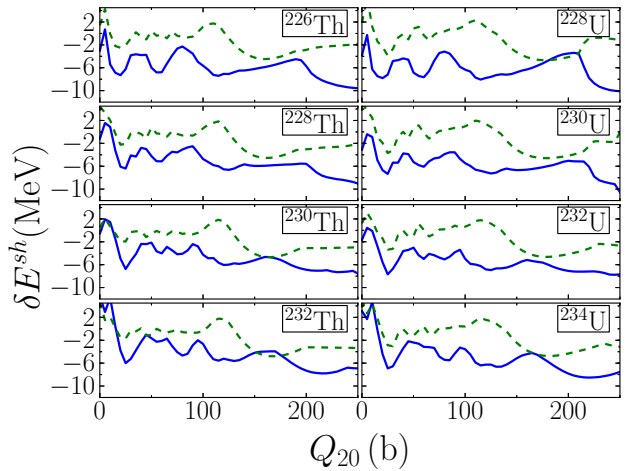}
 \caption[T]{\label{fig:SkM8ChainCompare}  (Color online) (solid lines) Neutron 
 and (dashed lines) proton
 shell correction energies as functions of $Q_{20}$ for (left) Th and (right) U even-even isotopes
 calculated in  SkM$^*$.
 }
\end{figure}

To study the dependence of this shell effect on neutron and proton  numbers, in Fig.~\ref{fig:SkM8ChainCompare} we plot
the individual  neutron and proton shell corrections in SkM$^*$ for Th and U isotopes as a function of $Q_{20}$. As expected, proton shell corrections weakly depend on the neutron number, and they all exhibit a minimum around $Q_{20}=160$\,b.
For $N=142$, the neutron shell correction shows two minima: one around
$Q_{20}=120$\,b and the second one around $Q_{20}=200$\,b. While the first minimum weakly depends on $N$, the second one is absent in $N=136, 138$ isotones. It is tempting, therefore, to associate the large neutron shell effect at $Q_{20}\approx 200$\,b with the prolate-deformed $N\approx 60$ fragments, and 
the large neutron shell effect at $Q_{20}\approx 120$\,b with the nearly-spherical $N\approx 54$ fragments. Since the maximum of the proton shell effects appears at the minimum of the neutron shell effect, the total shell correction is sensitive to both $N$ and $Q_{20}$. This cancellation helps to explain the shallow third minima obtained in DFT calculations.

In summary, we have found that our self-consistent SkM$^*$ and UNEDF1 models  predict  third minima for the $N=136, 138$  isotopes of Th and U.  How do the shell effects that favor third minima evolve with excitation energy?  In the next section, we turn to study PESs of $^{228, 232}$Th and $^{232}$U as a function of excitation energy, $E^*$.

\section{Excitation energy dependence}\label{section:th232pes}

\begin{figure}[htb]
 \includegraphics[width=0.8\columnwidth]{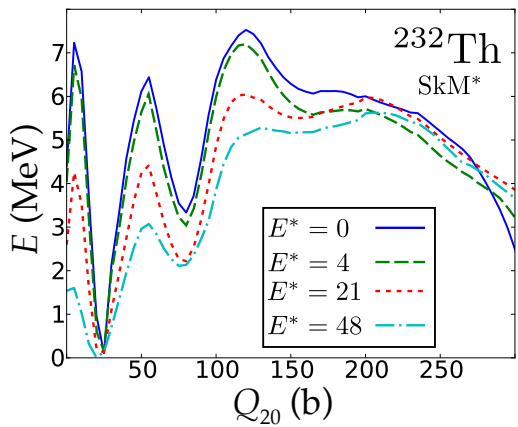}
 \caption[T]{\label{fig:th232Temperatures}
(Color online)  Isentropic potential-energy curves  for $^{232}$Th computed in SkM$^*$ at several values of excitation energy (in MeV). The minimum potential energy at given $E^*$
is normalized to zero in each case.}
\end{figure}

To discuss the excitation-energy dependence of the third minimum,
Fig.~\ref{fig:th232Temperatures} displays the SkM$^*$ potential-energy curves at constant entropy for $^{232}$Th at several excitation energies.  As excitation energy increases from $E^* = 0$ to $E^* = 48$\,MeV,  the second fission barrier is gradually reduced while a third barrier changes little.  This deepens the third-minimum pocket. 
The apparent stabilization of the third minimum at intermediate values of $E^*$
can be attributed to the interplay between pairing and shell effects 
\cite{RagnarssonNilsson,NazarewiczNPA1993}.  
Indeed, as discussed in, e.g., Ref. \cite{Egido2000,*Martin2003},  as the excitation energy increases, pairing correlations are quenched faster than the shell effects. This gives rise
to a reentrance of  shell effects  with $E^*$ in the third barrier region, so that the third minimum becomes more pronounced for moderate excitation energies.  Between $E^* \approx 21$\,MeV and $E^* \approx 48$\,MeV, the second barrier vanishes but the extended plateau around  the third minimum is still visible.  

\begin{figure*}[htb]
 \includegraphics[width=0.8\textwidth]{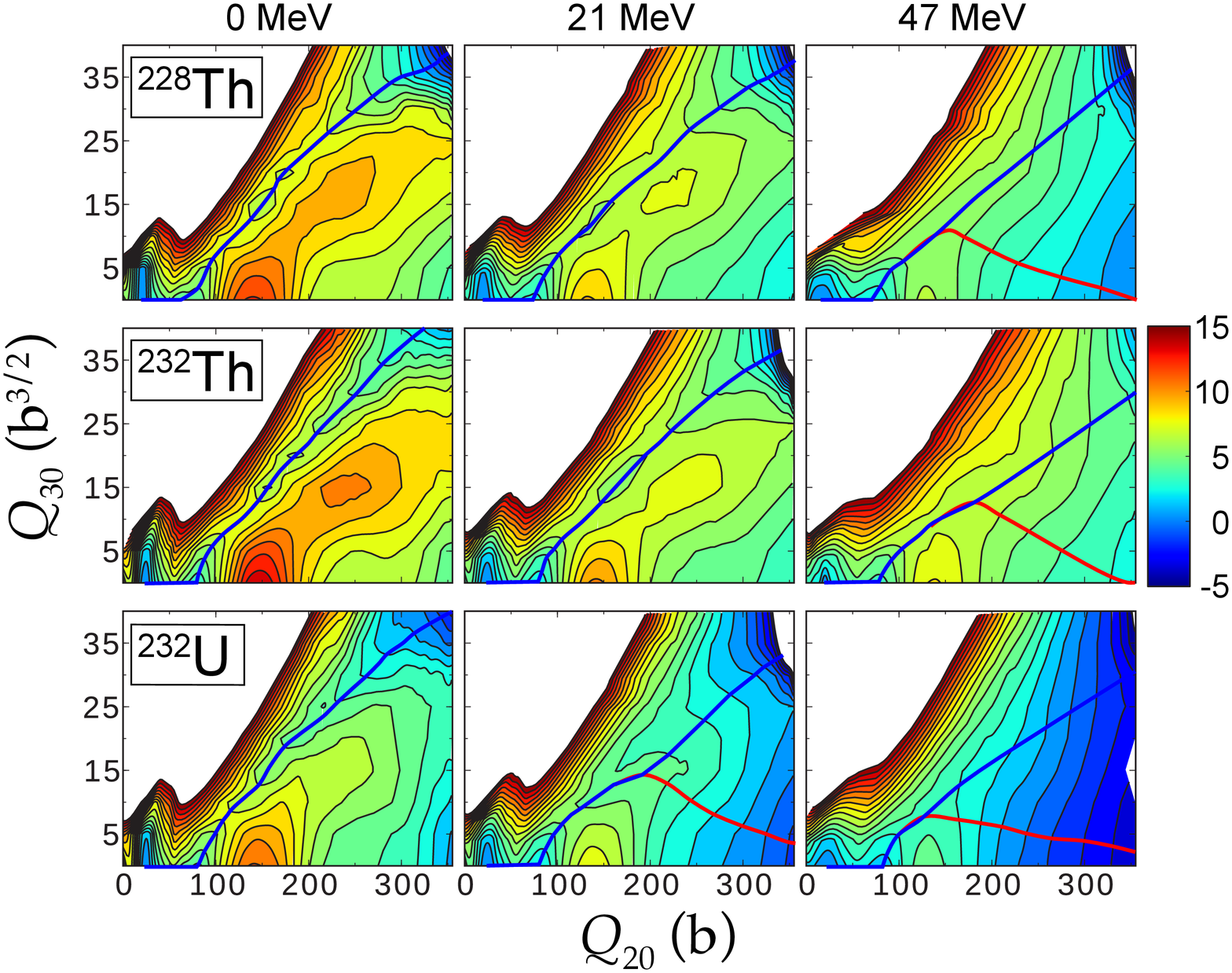}
 \caption[T]{\label{fig:FTcontours}
 (Color online) Isentropic potential-energy surfaces in the $(Q_{20}, Q_{30})$ plane for $^{228, 232}$Th and $^{232}$U calculated with SkM$^*$ at $E^*=0, 21$, and 47\,MeV.  The static  fission pathways are indicated. 
A constraint on the triaxial quadrupole moment $Q_{22}$ has been applied  to  minimize the total energy in the direction of $Q_{22}$.  Consequently,
the inner fission barriers are lowered due to the breaking of axial symmetry.  
 }
\end{figure*}
Because the presence of the third minimum at each excitation energy is sensitive to the accuracy with which the neighboring saddle points are found, we calculated two-dimensional PESs to assure us that the third minimum would not disappear when another degree of freedom is accounted for explicitly.  The two-dimensional PESs also enable us to assess whether the character of the dimolecular configuration at the third minimum changes with excitation energy.

We display the isentropic PESs in the $(Q_{20}, Q_{30})$ plane for $^{228, 232}$Th and $^{232}$U in Fig.~\ref{fig:FTcontours}. By constraining the triaxial moment $Q_{22}$, we  account for the effect of triaxiality on inner barriers. We trace the lowest-energy pathway from the ground state to the exit point of the barrier.  

Is there evidence that the dimolecular structure present at the third minimum for $E^* = 0$\,MeV persists as excitation energy increases?  As seen in Fig.~\ref{fig:FTcontours}, the third minimum in $^{228}$Th is actually rather robust -- a 1\,MeV pocket still remains at nearly the same collective coordinates (and nearly the same density profiles) at $E^* = 21$\,MeV, where pairing is completely quenched.  At higher excitation energies, the third minimum disappears as the symmetric fission pathway opens.  

The two-dimensional PES  shows the same evolution for $^{232}$Th as seen in the one-dimensional plots of Fig.~\ref{fig:th232Temperatures}. Namely,   as pairing is quenched around $E^*=21$\,MeV, the plateau around $Q_{20}=160$\,b  deepens into a well-developed third minimum. As with $^{228}$Th, at higher energies the third minimum disappears as  the symmetric fission channel opens.  

For $^{232}$U, however, our SkM$^*$ calculations do not appear to predict a clear third minimum anywhere in the range of $E^*$ studied.  A potential-energy shoulder appears at the lowest energies, so this may be the source of  the resonances observed in the experimental data presented in Ref.~\cite{CsigePRC80}.  

For all three isotopes, there is a strong preference for an asymmetric fission pathway that passes through the $(Q_{20}, Q_{30})$ coordinates of the third minima at low energies.  As $E^*$ increases, the barrier to symmetric fission lowers substantially so that the symmetric fission pathway gradually begins to compete with the asymmetric channel. 

This describes the situation seen experimentally: the mass distribution of  fission fragments in actinides is strongly asymmetric at low energies, and the symmetric mass yield increases with $E^*$.
For example, the experiment of Ref.~\cite{GuntherZPA295} measured the mass yield for the photofission of $^{232}$Th, reporting that the ratio of symmetric yield to asymmetric yield increases from $2$\% to $10$\% for a bremsstrahlung energy range (corresponding approximately to our excitation energy) of $15$ to $55$\,MeV.

\section{Conclusions}\label{section:th232conclusions}

This self-consistent FT-DFT study predicts very shallow third minima, or  shoulders, in the potential-energy surfaces of $^{232}$Th and $^{232}$U. 
In the lighter isotopes with $N=136$ and 138,  $^{226,228}$Th and $^{228,230}$U, the third minima are better developed. This can be traced back to  the neutron shell effect that reduces the third outer barrier at $Q_{20}\approx 200$\,b at $N=140$ and 142.
The shallowness, or absence,  of the third minimum in $^{232}$Th and $^{232}$U is a robust feature of many  DFT calculations, including  SkM$^*$, UNEDF1,  D1S, and HFB-14 models. We do not, therefore, confirm earlier MM predictions of deep hyperdeformed minima  in $^{232}$Th and $^{232}$U.

Our paper demonstrates  that the third minimum can be associated with a dimolecular configuration involving the spherical doubly magic $^{132}$Sn and a lighter Zr or Mo fragment in a deformed configuration. We show that the neutron shell effect that governs the existence of the third minimum and makes the third minimum more pronounced in $N=136$ isotopes as compared to $N=142$ systems is consistent with  the spherical-to-deformed shape transition in the Zr and Mo isotopes around $N=58$ ~\cite{Wood92,HeydeWood,Reinhard99}.
  
While the dimolecular structure persists through a range of excitation energy, the depth of the third minimum is found to be quite sensitive to excitation energy.  
Our FT-DFT  study predicts that third minima in Th isotopes become deeper at moderate excitation energies, where pairing correlations are quenched and shell effects become locally enhanced.  At large values of $E^*$, 
the conditions needed for the hyperdeformed metastable states  to exist deteriorate  as the symmetric fission channel  opens up.

While the inference of a hyperdeformed fission isomer from experimental data does rely on many  assumptions, the accumulated experimental evidence for the presence of resonances associated with reflection-asymmetric shapes is substantial and should not be considered lightly. Shallow  third minima (or shoulders) obtained in  self-consistent calculations are in fact consistent with the observed fission fragment distributions and resonances in fission probability. 
The absence of a well-developed local minimum in a static PES, a sole focus point of  Ref.~\cite{KowalPRC85}, does  not tell  the full story. 
Oftentimes, observed states can be  associated with configurations, which do not correspond to a minimum in  PES  \cite{Bennaz89,NazarewiczNPA1993} but are well separated from other states through the presence of specific quantum numbers. 
In this context, it would be a natural extension of this work to study the 
competition between the symmetric and asymmetric fission pathways, 
and clustering effects,
with a framework that accounts for fission dynamics, such as the generator coordinate method \cite{GouttePRC71,Dubray:2008lr,Bernard:2011qy} generalized to finite temperature.  
Also, the energetics of local minima, as well as diabatic configurations that may be associated with fission probability resonances, can be impacted by correlations associated with symmetry restoration, such as those discussed in Refs.~\cite{Tajima1993409,Bender04}.  
We see some early evidence of this impact in our UNEDF1 calculations employing the approximate number projection, in which third minima are generally shallower and even disappear for all but the lightest isotopes.  Isolating the effects due to particle number projection would be an interesting topic for future study.  
An additional topic for future study consists in following whether the dimolecular configuration persists from the third minimum to scission. To this end, one could apply the techniques of Ref.~\cite{YounesPRL107}.  In light of the recent experiment on $^{238}$U \cite{Csige:2013aa}, it would be particularly interesting to extend this study to heavier isotopes of uranium and thorium.  

\begin{acknowledgments}
Useful discussions with A. Staszczak, N. Schunck, and M. Warda
are gratefully acknowledged. This work was
supported by
the U.S. Department of Energy under
Contracts No.\ DE-FG02-96ER40963
 (University of Tennessee), No.\ DE-FG52-09NA29461 (the Stewardship Science Academic Alliances program),  DE-AC07-05ID14517 (NEUP grant subaward 00091100), and No.\
DE-SC0008499    (NUCLEI SciDAC Collaboration).  An award of computer time was provided by the Innovative and Novel Computational Impact on Theory and Experiment (INCITE) program. This research used resources of the Oak Ridge Leadership Computing Facility located in the Oak Ridge National Laboratory, which is supported by the Office of Science of the Department of Energy under Contract DE-AC05-00OR22725.  This work was performed under the auspices of the U.S. Department of Energy by the Lawrence Livermore National Laboratory under Contract No. DE-AC52- 07NA27344.
\end{acknowledgments}

\bibliographystyle{apsrev4-1}
\bibliography{jmcdonnell_th232}

\end{document}